\documentclass[11pt]{article}
\usepackage[margin=1.3in]{geometry}
\usepackage{amsfonts}
\usepackage{amssymb}
\usepackage{setspace}
\usepackage{hyperref}
\usepackage{enumitem}
\usepackage{natbib}
\usepackage{comment}

\title{{\textbf{The \emph{Kal$\overline{\textrm{a}}$m} Cosmological Argument Meets The Mentaculus}}}
\author{{\normalsize \textbf{Dan Linford}}}
\date{}

\usepackage[T1]{fontenc}

\begin{document}

\maketitle


\begin{abstract}
According to the orthodox interpretation of bounce cosmologies, the universe was born from an entropy reducing phase in a previous universe. To defend the thesis that the whole of physical reality was caused to exist a finite time ago, William Lane Craig and co-author James Sinclair have argued the low entropy interface between universes should instead be understood as the beginning of two universes. Here, I present Craig and Sinclair with a dilemma. On the one hand, if the direction of time is reducible, as friends of the Mentaculus -- e.g., David Albert, Barry Loewer, and David Papineau --  maintain, then there is reason to think that the direction of time and the entropic arrow of time align. But on that account, efficient causation is likely reducible to non-causal phenomena. In consequence, contrary to Craig and Sinclair's theological aims, things can begin to exist without causes. On the other hand, if the direction of time is not reducible, Craig and Sinclair's interpretation of bounce cosmologies is unjustified. Lastly, a reply to a potential objection motivates a discussion of how to interpret bounce cosmologies on the tensed theory of absolute time favored by Craig and Sinclair. I offer two interpretations of bounce cosmologies that, given a tensed theory of absolute time, are preferable to those Craig and Sinclair offer, yet inconsistent with their project in natural theology; on one interpretation, the universe does not require a supernatural cause and, on the other, bounce cosmologies represent the universe as never having begun to exist.
\end{abstract}


\tableofcontents



\doublespacing

\section{Introduction}

The universe began in a special state, with a stupendously and surprisingly small value of the universe's entropy, approximately fourteen billion years ago. According to the orthodox interpretation of one family of cosmological models -- that I will refer to as `bounce cosmologies' -- the low entropy of our universe's beginning can be explained by an entropy reducing phase in a previous universe. Nonetheless, William Lane Craig, and his sometimes co-author James Sinclair, have brought the orthodox interpretation into question. They defend the \emph{Kal$\overline{\textrm{a}}$m} Cosmological Argument for Theism (\cite{CraigSinclair:2009, CraigSinclair:2012}):

\begin{enumerate}
    \item Everything which begins to exist has a cause for its existence.
    \item \label{KCA2}The universe began to exist.
    \item Therefore, the universe has a cause for its existence.
\end{enumerate}

Proponents of the \emph{Kal$\overline{\textrm{a}}$m} argument go on to argue that God is the only plausible candidate for the cause of the universe's existence. Nonetheless, on the orthodox interpretation, if a bounce cosmology were shown to be a good approximation of the universe we inhabit, we would have an alternative explanation for the universe's beginning to exist, that is, the physical states in the entropy reducing phase prior to our universe. Craig and Sinclair have argued that the orthodox interpretation is incorrect; in their view, the direction of time is indicated by the direction of entropy increase. So, instead of interpreting the interface between the two universes as the death of one universe and the birth of another, Craig and Sinclair interpret the interface as the birth of two universes (\cite{CraigSinclair:2009}, \cite{CraigSinclair:2012}; \cite{CarrollCraig:2016}).

In this paper, I show that either Craig and Sinclair's interpretation is inconsistent with the \emph{Kal$\overline{\textrm{a}}$m} argument or inconsistent with their broader theological project. To do so, I present Craig and Sinclair with what I call the `Interface Dilemma': either the macrophysically observable direction of time is reducible or irreducible. If the direction of time is reducible, then, even if the interface should be interpreted as an absolute beginning for two universes, the first premise of the \emph{Kal$\overline{\textrm{a}}$m} argument is probably false. This is because on the most promising reduction of the direction of time -- that is, the \emph{Mentaculus}, as defended by David Albert [\cite{albert_2000, Albert:2015}] and  Barry Loewer [\citeyear{Loewer:2012}; \citeyear{Loewer:2012b}; \citeyear{Loewer:2020}] -- efficient causation is most likely reducible. If efficient causation is reducible, then things that begin to exist can do so without an efficient cause.\footnote{Sean Carroll has previously advanced a similar argument. In Carroll's view, two features allow us to construct causal explanations for objects that arise \emph{within} physical reality: first, that the objects obey the laws of physics and, second, that there is a low entropy boundary condition in the past. When we consider the totality of physical reality, we are no longer considering objects within physical reality. There is no collection of physical laws that apply in that broader context and there is no low entropy boundary condition beyond the physical world. Therefore, Carroll argues, we have no `right to demand some kind of external cause' \cite[pp. 67-8]{CarrollCraig:2016}; also see \cite{Carroll:2005, Carroll:2012}. Carrroll does not develop his argument to the extent that I develop the first arm of the Interface Dilemma nor does Carroll consider the issues that I raise in the second arm. Moreover, my argument does not rely on the thesis that physical laws apply only to objects within physical reality.} On the other hand, if the direction of time is not reducible, then Craig and Sinclair's interpretation of bounce cosmologies is unjustified. Lastly, a reply to a potential objection to my dilemma will motivate a discussion of Craig and Sinclair's brand of the tensed theory of time. As I will show, their tensed theory of time affords two interpretations of bounce cosmologies that are preferable to the interpretations Craig and Sinclair have offered. In one of the interpretations that I offer, the universe began to exist, but without need for an external, supernatural cause. If so, theistic conclusions cannot be drawn from the \emph{Kal$\overline{\textrm{a}}$m} argument. The other interpretation that I offer is inconsistent with the second premise of the \emph{Kal$\overline{\textrm{a}}$m} argument; according to this interpretation, bounce cosmologies represent physical reality as never having begun to exist.

\section{\label{First_Arm}The first arm: the direction of time is reducible}

As discussed in the introduction, the Interface Dilemma has two arms. On the first arm, there's the view that the direction of time is reducible, with the consequence that the first premise of the \emph{Kal$\overline{\textrm{a}}$m} argument (that everything that begins to exist has a cause for its existence) is probably false. On the first arm, we have the first two premises and subconclusion of the Interface Dilemma:

\begin{enumerate}
    \item If the direction of time is reducible then, probably, not all events have efficient causes.\label{if_reducible_then_not_all} 
    \item If, probably, not all events have efficient causes then the first premise of the \emph{Kal$\overline{\textrm{a}}$m} argument is probably false.\label{if_not_all_then_P1_false} 
    \item So, if the direction of time is reducible then the first premise of the \emph{Kal$\overline{\textrm{a}}$m} argument is probably false (hypothetical syllogism from \ref{if_reducible_then_not_all}, \ref{if_not_all_then_P1_false}).\label{subconc1} 
\end{enumerate}

What does it mean to say that the direction of time is reducible and why think that, if the direction of time is reducible, efficient causation is reducible to non-causal phenomena? The most compelling view on which the direction of time is reducible has been provided by David Albert ([\citeyear{albert_2000}], [\citeyear{Albert:2015}]) and Barry Loewer ([\citeyear{Loewer_2007}], [\citeyear{Loewer:2012}], [\citeyear{Loewer:2012b}], [\citeyear{Loewer:2020}]) and, together with David Papineau ([\citeyear{papineau_2013}]), they have argued that their reductive programme has the consequence that efficient causation is reducible. Herein, I will refer to this programme as the `Albert-Loewer-Papineau reductive programme', or ALP.\footnote{Ludwig Boltzmann similarly pursued a reductive explanation for the direction of time in the nineteenth century; see, for example, \cite{Steckline:1983}.} I need to set two key elements of ALP on to the table. First, I need to sketch the sense in which, according to ALP, the direction of time is reducible. Second, I need to sketch why, given the reducibility of the direction of time, ALP maintains the reducibility of efficient causation.

\subsection{ALP on the direction of time}

With some qualifications to be introduced momentarily, the direction of time does not appear in our best microphysical theories;\footnote{Some care should be taken in distinguishing microphysical descriptions from macrophysical descriptions. See \cite[p. 134]{papineau_2013}, \cite[p. 148]{Ney:2016}. What current physical theory takes to be fundamental is probably the result of coarse graining still more fundamental degrees of freedom that will appear in a successor theory. For present purposes, it suffices that we can distinguish phenomena that are fundamental \emph{according to our best physical theories} from phenomena that are not fundamental.} that is, microphysical phenomena are best described in a way that does not distinguish the past from the future. To see what this means, we can consider a world whose microphysics is described by Newtonian mechanics. This description will not be correct for our world, in which the microphysics is instead -- as far as anyone can tell so far -- best described by quantum field theory. Nonetheless, both Newtonian mechanics and quantum field theory are temporally symmetric theories.\footnote{\label{t-asymmetry-footnote}An anonymous reviewer worried that I've moved too quickly. Perhaps a solution to the measurement problem will reassert time asymmetry, so perhaps quantum field theory will turn out to be time asymmetric after all. In reply, objective wavefunction collapse is not a part of textbook quantum field theory and I don't find objective collapse theories promising. Nonetheless, there are other reasons to be worried about the time asymmetry of quantum field theory; for example, one might worry about the observed violation of $CP$ symmetry, as I discuss elsewhere in this paper. The important point for my present purposes is that the distinction between the past and the future made in fundamental physics (if fundamental physics really does distinguish the past from the future) are unlikely to explain the distinction between causes and their effects or any of the other macrophysically observable temporal asymmetries.} Consider a ball in a vacuum traveling at a fixed speed in a fixed direction. Suppose that the ball impacts and rebounds off of a wall. I will suppose that the wall is not fixed to the floor but, instead, rests on frictionless rollers. When the ball rebounds off the wall, to conserve momentum, the wall must begin moving. (I am going to suppose that neither the ball nor the wall heat up when the ball rebounds off of the wall. We can safely make this assumption because we've restricted ourselves to Newtonian mechanics, which does not include a theory of heat.) We have three events: (1) the ball is traveling at a fixed velocity, (2) the ball rebounds off the wall, and (3) both the ball and the wall are traveling at fixed velocities. If we run the sequence of events in reverse, we find (1*) the ball and the wall are traveling at fixed velocities, (2*) the ball and the wall collide  with each other, and (3*) the wall is at rest, but the ball continues to move. Both sequences of events describe solutions to the equations of Newtonian mechanics. As far as Newtonian mechanics is concerned, the universe runs equally well in reverse as the universe runs forward.

What this temporal symmetry amounts to -- particularly in the case of quantum field theory -- turns out to be a delicate and subtle matter. The equations of fundamental physics are not invariant under the simple operation of replacing every instance of $t$ with an instance of $-t$. According to the standard argument in textbooks, one can define an effective time reversal operation that takes some sequence of physical states $S \equiv \{S_1$, $S_2$, ..., $S_N\}$ and returns a corresponding sequence of states $S^* \equiv \{S_1^*$, $S_2^*$, ..., $S_N^*\}$ that are said to be the `time reversal' of $S$. To say that physics is `time reversal invariant' is then to say that the time reversal operation returns a nomologically permissible sequence of events.\footnote{Albert disagrees with this definition of `time reversal', but agrees that whatever irreversibility might obtain in the microphysics, for example, in the weak interaction's violation of $T$ symmetry, cannot explain the irreversibility of macrophysics needed for explaining the asymmetry of causation \cite[p. 21]{albert_2000}; for a reply, see  \cite{Earman_2002}. Also see footnote \ref{t-asymmetry-footnote}.} In any case, this sort of operation cannot be defined for the causal relations observed in the special sciences. For example, every morning, I see eggs frying but I have never seen eggs \emph{unfrying}. So how do causal relations get into the world? ALP maintains that the macrophysical direction of time -- and so, as we will see, the asymmetry of efficient causation -- is the result of some sort of averaging over, and coarse graining of, microphysical phenomena (\cite{albert_2000}, \cite{Albert:2015}; \cite{Loewer:2012}, \cite{Loewer:2012b}, \cite{Loewer:2020}). And this coarse graining can -- if ALP's programme succeeds -- be parlayed into an explanation of macrophysical efficient causation. 

Phase space is the space of possible microphysical states of a system.  From a macrophysical perspective, in which the microphysics has been averaged over, information about the exact microphysical configuration is not available; consequently, one can only identify a region of phase space to which the universe's exact microphysical state belongs. But given that the universe occupies a sufficiently small phase space region at some time $T$, one can predict, with overwhelming probability, that the universe will evolve at time $T+1$ to a state that lives in a larger phase space region. And this is so even though the microphysical dynamics describing the universe's precise trajectory through phase space are time symmetric.

The size of the phase space region -- that is, the `number' of microphysical configurations\footnote{That is, the size of the region given by the appropriate measure on phase space and not the cardinality of the collection of states in that region.} -- consistent with the macrophysical description is termed the \emph{entropy}.\footnote{The entropy is defined as the sum -- or the integral in the case of continuous systems -- of $\rho_i \log(\rho_i)$, where $\rho_i$ is the probability associated with the $i$th microphysical state of the system. This definition is provably equivalent to the size of a phase space region provided that we assume the uniform measure over phase space, that is, in Newtonian mechanics, the Lebesgue measure.} The amount of entropy characterizing a system has observable macrophysical consequences. To see this, consider a crowd trying to push a boulder. Merely commanding the individuals comprising the crowd to run at the boulder is a poor strategy; without coordination between the individuals, the boulder will, at best, `quiver' when, by chance, multiple individuals impact on one side or the other. If we instead command all of the individuals comprising the crowd to charge at the boulder from a specific angle -- that is, if we command the individuals in a way that coordinates their efforts -- then the crowd can collectively do work to move the boulder. Notice that there are many more ways for the crowd to charge the boulder in a disorganized fashion than there are ways for the crowd to charge the boulder in a coordinated fashion. Likewise, if we want a gas to do work in moving a piston, we need for the motions of the gas atoms to be coordinated in the right sort of way, and coordinations like that occupy a small phase space volume. When systems move towards equilibrium -- that is, when their microphysical parts become successively more disorganized -- we can extract successively less work from the system. In the early universe, there must have been a tremendous amount of energy available for doing work during the universe's subsequent evolution. This suggests that the universe began with much less entropy than the universe presently has.

There is a more significant reason for thinking that the universe began in a lower entropy configuration. Consider, for example, our records of the past. Our records of the past require a three part relation between the present moment, the moment the record was created, and the moment before the record was created. There are a stupendously large number of trajectories through phase space along which record impostures indistinguishable from the records that we possess merely fluctuated into existence. That is, there are a stupendously large number of trajectories from higher entropy regions that pass through the region of phase space containing record impostures macrophysically indistinguishable from our records. So, to ensure that the record that we possess did not merely fluctuate into existence, we need to ensure that our recording devices, in the moment before our records are created, were in the appropriate state waiting to create the record, that is, the ready state. But to ensure that the device was in the device's ready state requires a record of the device having been in the ready state. We are off to the races on a regress of successive ready states, until we come to the ultimate ready state, that is, a low entropy constraint for the beginning of the universe \cite[p. 118]{albert_2000}. For ALP, this constraint is codified in a hypothesis they call the `Mentaculus', consisting of the conjunction of three principles. First, whatever the fundamental dynamical laws happen to be. Second, the Past Hypothesis, that is, the hypothesis that the universe began in the low entropy macrophysical state $M(0)$.\footnote{$M(t)$ is the universe's macrophysical state at time $t$, so that $M(0)$ is the state at $t=0$.} And, third, the Statistical Postulate, that is, the specification of a uniform probability measure over the portion of phase space consistent with whatever information we happen to have about the physical world \cite[p. 96]{albert_2000}.

According to ALP, every formalizable proposition $p$ concerning the state of the physical world -- and so nearly every statement that could ever be made in the sciences -- is assigned an objective probability by the Mentaculus, that is, $Pr(p|L\&M(0)\&M(t))$, where $L$ denotes the fundamental laws. For this reason, Albert and Loewer have called the Mentaculus the `probability map of the world'. According to ALP, every macrophysical temporal asymmetry can be (in principle) recovered from the Mentaculus as a consequence of the fact that the universe began in a low entropy configuration, despite the fact that the fundamental dynamical laws are time reversal invariant. For example, the probability that an egg in a hot frying pan will fry, given the universe's low entropy initial state and the present environment of the egg, is close to $1$, while the probability that the egg will unfry is close to zero.

\subsection{ALP on efficient causation}

The project of explaining all temporal asymmetry -- including the asymmetry of efficient causation -- in terms of the Mentaculus is ongoing. However, there are several reasons for thinking that the Mentaculus project, if successful, will provide a reduction of efficient causation. Efficient causation is a time asymmetric phenomenon. Causes do not typically, if ever, occur after their effects.\footnote{An anonymous reviewer comments that this statement assumes a globally hyperbolic relativistic space-time. If we allow for the possibility of closed time-like curves, then causes may occur after their effects. Nonetheless, my statement that causes do not typically occur after their effects would still be true if closed time-like curves are rare. My statement may require a different and (albeit small) revision within the view that efficient causation is reducible. According to ALP, unfrying eggs is stupendously improbable but not impossible. By a parallel set of considerations, on the view that efficient causation is reducible, effects may precede causes, but only with an unimaginably low probability. For discussion, see chapter two in \cite{Albert:2015}.} ALP maintains that all macrophysical time asymmetry will be given a reductive explanation in terms of the Mentaculus and, therefore, maintains that efficient causation, qua macrophysical time asymmetry, will be given a reductive explanation in terms of the Mentaculus. As Alyssa Ney puts the point, `from the point of view of microphysics, given an individual event, there is no objective distinction between which events make up that event’s past and which its future. Therefore, there is no microphysical distinction between which are its causes and which its effects. Thus, there are no facts about microphysical causation' \cite[p. 146]{Ney:2016}. 

Promisingly, steps have already been taken to show how to recover the apparent openness of the future as well as the truth conditions for causally relevant counterfactuals from the Mentaculus. Consider a collection of billiard balls on a table. If some specific ball was moving one minute ago and is not moving now, then the ball must have been involved in a collision at some time between now and one minute ago. That is, that the impact happened counterfactually depends on the present state of the ball. However, whether the ball will be moving one minute in the future counterfactually depends upon nothing less than the motions of all of the balls \cite[pp. 127-8]{albert_2000}. For ALP, the difference in the counterfactual dependence between present macrophysical states and past or future macrophysical states is explained by the temporal asymmetry provided for the macrophysical world by the Mentaculus. If, on the supposition that microphysical causation exists, one expects to find microphysical causation in physics, then one has reason for rejecting microphysical causation. In the microphysical world, events merely happen.

ALP is distinct from the sort of causal skepticism often attributed to David Hume and associated with opposition to metaphysics, according to which efficient causation is a mere projection of the human mind and, consequently, not a real feature of the world. In contrast, ALP is not anti-metaphysical; for Albert, Loewer, and Papineau, we can and should read our metaphysics off of our best science. ALP denies microphysical causation because, in the metaphysical view that they think we should read off of our best science, causation is not one of the features of microphysical phenomena. Nonetheless, reductionists are not eliminitivists. Just as temperature is a real feature of macrophysical gases that does not apply to the microphysical constituents of gases, according to ALP, efficient causation is a real feature relating macrophysical events (see, for example, \cite[p. 297]{Loewer_2007}). In other words, to borrow a phrase from Daniel Dennett, efficient causation is a real pattern \cite{Dennett:1991}. And we should be ontologically committed to efficient causation because efficient causation is an ineliminable feature of our best explanations in the special sciences (\cite{Loewer:2012}, \cite{Loewer:2020}; \cite{papineau_2013}). Likewise, we should distinguish ALP from the view that quantum indeterminism demonstrates that efficient causation is inapplicable to microphysical processes. Our example of the Newtonian world demonstrated ALP's consistency with determinism.

\subsection{Synthesis}

Having set ALP's reduction of the direction of time, and the consequent causal reductionism, on the table, I now turn to synthesizing these elements into the first arm of the Interface Dilemma. The first premise of the \emph{Kal$\overline{\textrm{a}}$m} argument -- that everything which begins to exist has a cause of its existence -- should be interpreted to mean that all things that begin to exist have an efficient cause for their existence. Hume's causal skepticism has been thought to have negative consequences for the cosmological argument (for example, \cite[pp. 113-28]{russell:2008}; \cite{Russell:2013}). Hume's causal skepticism is distinct from and logically independent of causal reductionism; for example, Hume's causal skepticism applies to all efficient causation, whether macrophysical or microphysical. In contrast, causal reductionism is consistent with macrophysical efficient causation in the special sciences or in ordinary life, while denying efficient causal relations in fundamental physics. So, if efficient causation is reducible -- that is, if ALP is true -- then the first premise of the \emph{Kal$\overline{\textrm{a}}$m} argument is false because microphysical events generally lack efficient causes. And if events do not generally require efficient causes, then the beginning of the universe would not require an efficient cause, God or otherwise.\footnote{A reviewer objects that Craig and Sinclair could restrict the \emph{Kal$\overline{\textrm{a}}$m} argument to macrophysical phenomena and that the beginning of the universe could be understood as a macrophysical phenomenon. Perhaps the universe has a macrophysical cause for its beginning to exist? I have difficulty seeing how such a reformulation of the \emph{Kal$\overline{\textrm{a}}$m} argument would allow Craig and Sinclair to overcome the difficulty posed by ALP's causal reductionism. On the one hand, supposing that the \emph{Kal$\overline{\textrm{a}}$m} argument could be saved in this way and that one could show that the universe had a macrophysical cause, this would be a merely pyrrhic victory for Craig and Sinclair. After all, if causation is a reducible relation obtaining between macrophysical entities, then one may doubt that God, qua non-physical entity, could be the efficient cause of the universe. On the other hand, I have a difficult time seeing how the \emph{Kal$\overline{\textrm{a}}$m} argument could be saved. If physics has shown us that we need to give up irreducible causation and offered a new set of explanatory criteria for fundamental explanations, then we have little reason to think that efficient causation is applicable to a situation as exotic as the birth of physical reality.} ALP is the most promising view according to which the direction of time is reducible; thus, if the direction of time time is reducible, then, probably, ALP is true, and the first premise of the \emph{Kal$\overline{\textrm{a}}$m} argument is probably false.

\subsection{An objection to premise \ref{if_not_all_then_P1_false}}

Now recall the first arm of the dilemma as presented at the outset of section \ref{First_Arm}:

\begin{enumerate}
    \item If the direction of time is reducible then, probably, not all events have efficient causes. 
    \item If, probably, not all events have efficient causes then the first premise of the \emph{Kal$\overline{\textrm{a}}$m} argument is probably false. 
    \item So, if the direction of time is reducible then the first premise of the \emph{Kal$\overline{\textrm{a}}$m} argument is probably false (hypothetical syllogism from \ref{if_reducible_then_not_all}, \ref{if_not_all_then_P1_false}). 
\end{enumerate}

Because the first arm is a valid argument, objections must show either that at least one of the two premises has not been adequately supported, that at least one premise is probably false, or that the conjunction of the two premises is more probably false than true. Premise \ref{if_reducible_then_not_all} follows from the view that ALP is our best theory of the reducibility of macrophysical temporal directedness. So, objections would have to be made to \ref{if_not_all_then_P1_false}.

Quentin Smith has previously targeted the first premise of the \emph{Kal$\overline{\textrm{a}}$m} argument -- that everything which begins to exist has a cause for its existence, herein, the causal principle -- by drawing upon physics [\citeyear[p. 121]{Smith_1993_Uncaused}]. Though Smith's argument draws from different parts of physics than my argument, perhaps Craig can construct a parallel reply to the one that he has offered to Smith's argument. According to Smith, there are chancy quantum mechanical events -- vacuum fluctuations -- that, he argues, lack efficient causes because they occur without temporally prior events that necessitate them. Craig replied that though there are no temporally prior events that necessitate any given vacuum fluctuation, there are antecedent physical conditions necessary for the occurrence of vacuum fluctuations. In Craig's view, the universe beginning to exist uncaused would require an event with no physically necessary antecedents. As he writes, `The appearance of a particle in a quantum vacuum may thus be said to be spontaneous, but cannot properly be said to be absolutely uncaused, since it has many physically necessary conditions. To be uncaused in the relevant sense of an absolute beginning, an existent must lack any non-logical [sic] necessary or sufficient conditions whatsoever' \cite[pp. 146-7]{Craig_1993_Caused}. That is, according to Craig, vacuum fluctuations have causes in the sense that there are explanatorily prior and physically necessary conditions that precede the vacuum fluctuations. In a parallel reply to my argument, Craig could concede that, in one sense of `causation', causation is reducible, so that not all events have efficient causes, while maintaining that, in a second sense of `causation', causation is not reducible. In the second sense, nothing begins to exist without the obtainment of specific explanatorily prior physically necessary conditions.

This objection does not work. Supposing that ALP is true, so that efficient causation is reducible, there may be explanatorily prior, physically necessary conditions for the coming into being of some microphysical entity. But that the conditions are explanatorily prior does not require any particular temporal relationship. Recall Ney's summary of ALP: `from the point of view of microphysics, given an individual event, there is no objective distinction between which events make up that event’s past and which its future'. Even if the coming into being of $E$ requires explanatorily prior, physically necessary conditions $C$, provided that ALP is true, so that, microphysically, there is no objective way to distinguish past and future directions, the explanatorily prior, physically necessary conditions need not fall in any particular temporal direction with respect to $E$. In other words, ALP allows us to maintain that the explanatorily prior and physically necessary conditions for the universe's `beginning' can fall in the temporal direction away from the beginning.

Moreover, whereas Craig has argued that God is both explanatorily prior to and simultaneous with the beginning of the universe, ALP allows us to maintain that entities do not require explanatorily prior or simultaneous causes for their coming into being. On the General Relativistic description, when a particle traverses a geodesic passing into a curvature singularity, the particle ceases to exist at the singularity. If this process happens microphysically, then ALP tells us that the process is not objectively distinguishable from a process in which a particle emerges from a curvature singularity. In the former case, there is nothing explanatorily posterior to or simultaneous with the singularity that `uncreates' the particle. So, in the time reverse process, nothing explanatorily prior to or simultaneous with the singularity would create the particle. According to ALP, both are descriptions of the same state of affairs. Consequently, if ALP is true, then microphysical entities do not require causes, even in Craig's more general sense, for their coming into being.

According to another reply that Craig offers to Smith, we should endorse the causal principle because we have a strong a priori intuition that `something cannot come out of nothing'. If so, perhaps an `inductive survey of existents in spacetime' cannot demonstrate the falsehood of the causal principle \cite[p. 147]{Craig_1993_Caused}. For this reason, Craig might argue that the empirical support for ALP cannot demonstrate that the causal principle is probably false. I don't share Craig's confidence in our intuitions concerning the universe as a whole or what we should expect in the far flung depths of nature. We've already seen -- in the thought experiment in the previous paragraph -- that if ALP turns out to be right, our intuition would have been shown to be incorrect.

In any case, ALP can explain why we have the strong intuition that the causal principle is true. For ALP, the intuition was hard-wired into us through a selection history in which our ancestors interacted with macrophysical events, each of which could be expected to have an identifiable efficient cause.\footnote{John Norton has similarly argued that efficient causation is a folk scientific notion that does not survive into mature science \cite{Norton:2003}. Norton's eliminitivism is more radical than ALP's reductionism.} Our intuition should be taken to range over macrophysical phenomena -- for which events do have causes -- but should not be taken to range over microphysical phenomena -- for which, according to ALP, events do not have causes. Since early modernity, the explanatory categories that we utilize when we plumb nature's depths have been repeatedly revised and have been shown to be far removed from the categories relevant for explanations in the manifest image or in the special sciences.\footnote{For a wonderful discussion of this point, see \cite[pp. 146-7]{Waismann:1961}.} With Galileo, we had to give up the Aristotelian demand that uniform motion requires a cause and some have argued that quantum mechanics forces us to give up a demand for deterministic causation. Perhaps physics has simultaneously shown us that we need to give up irreducible causation and offered a new set of explanatory criteria for fundamental explanations. If physics has shown us that we need a different collection of explanatory criteria for the far flung depths of nature than are required for the special sciences, then, contrary to theological ambitions, we should not expect the explanatory apparatus of the special sciences to return in fundamental metaphysics.

\section{The second arm: the direction of time is not reducible}

On the other hand, if the direction of time is not reducible, then the direction of efficient causation need not align with the direction of the entropic arrow of time. In this case, two negative consequences follow for Craig and Sinclair's interpretation of bounce cosmologies. This provides us with the next three premises and another sub-conclusion of the Interface Dilemma:

\begin{enumerate}[resume]
    \item If the direction of time is not reducible then the direction of time does not necessarily correspond to the direction of the entropic arrow of time.\label{if_not_reducible_then_not_cor} 
    \item If the direction of time does not necessarily correspond to the direction of the entropic arrow of time then\label{premiseAandB}
    \begin{enumerate}
        \item the interpretation of the interface as a beginning without a physical cause is unjustified and \label{beginning_unjustified}
        \item the claim that events on one side of the interface cannot be the efficient causes of events on the other side of the interface is unjustified.\label{eff_cause_unjustified} 
    \end{enumerate}
    \item If (\ref{beginning_unjustified}) and (\ref{eff_cause_unjustified}) then Craig and Sinclair's interpretation of bounce cosmologies is unjustified.\label{If_ab_then_unjustified} 
    \item So, if the direction of time is reducible then Craig and Sinclair's interpretation of bounce cosmologies is unjustified (hypothetical syllogism from \ref{if_not_reducible_then_not_cor}--\ref{If_ab_then_unjustified}).\label{subconc2} 
\end{enumerate}

To explicate this subargument, I turn to unpacking Craig and Sinclair's interpretation of bounce cosmologies. With the discovery of General Relativity in the early twentieth century, for the first time, physicists possessed a set of equations -- the Einstein Field Equations (EFE) -- whose solutions are possible space-times. On the assumption that the universe is homogeneous, isotropic, and with some constraints on the universe's energy content -- that is, assumptions once thought to be at least approximately true on cosmological length and time scales -- the EFE predict that our space-time cannot be continued indefinitely far into the past. Instead, one comes to a temporal boundary, in the sense that, prior to any moment, there is only a finite amount of time.

The boundary comes in the form of a curvature singularity where space-time becomes mathematically undefined. Theorems from Hawking and Penrose showed that this result does not depend upon the unrealistic assumption that the universe is isotropic or homogeneous.\footnote{Hawking and Penrose assume conditions on the energy contents of the universe that can be violated by quantum fields. For the history of singularity theorems up through the Hawking-Penrose theorems, see \cite{Earman:1999}.} Subsequently, Arvind Borde, Alan Guth, and Alex Vilenkin developed a new and more general theorem, the BGV theorem \cite{Borde:2003}. According to the BGV theorem, any geodesic along which the average of the universe's expansion rate would be measured to be greater than zero must terminate in the past, thereby suggesting that classical expanding space-times are singular. Borde, Guth, and Vilenkin interpreted their theorem to show where our understanding of space-time runs out. Craig and Sinclair have instead interpreted the theorem to show that space-times that are, on average, expanding must have a temporal boundary in the finite past. (As they write, `If the universe (or multiverse) expands (on average), then it has a beginning, period', \cite[p. 108]{CraigSinclair:2012}.)\footnote{Craig and Sinclair have mischaracterized what the BGV theorem has shown. First, contrary to their claim that the BGV theorem shows the universe could not have been expanding forever, the BGV theorem does not apply to non-classical space-times. To be sure, the BGV theorem is not restricted to the space-times that are the solutions to the EFE. Nonetheless, one can identify non-classical space-times to which the BGV theorem does not apply and in which the universe could have been expanding from eternity past. Second, while the BGV theorem shows that, in classical space-times, geodesics, along which one would measure a positive, non-zero average expansion rate, cannot be continued to eternity past, the BGV theorem does not tell us that all geodesics cannot be continued to eternity past. As Guth has pointed out, the BGV theorem provides no upper bound to the lengths of all of the geodesics within the space-times to which the theorem applies \cite[p. 6623]{Guth2007}. Andrei Linde writes that, `If this upper bound does not exist, then eternal inflation is eternal not only in the future but also in the past' \cite[p. 17]{Linde:2008}. Third, the BGV theorem does not even tell us that all such geodesics must meet up in a common singularity. As Linde describes, `at present we do not have any reason to believe that there was a single beginning of the evolution of the whole universe at some moment $t = 0$' \cite{Linde:2008}. Instead, the BGV theorem is consistent with the construction of a space-time in which given a geodesic $\gamma_1$ that terminates at some proper time $T_1$ in the past, there will exist another geodesic $\gamma_2$ that terminates at some proper time $T_2$ even further in the past. In a space-time like that, I have difficulty understanding what the claim that the universe began to exist could mean.} 

One can avoid the cosmological curvature singularity by stipulating that the universe's average expansion rate is less than or equal to zero or that we do not inhabit a classical space-time. For example, the universe could be cyclic (as in, for example, \cite{Ijjas_2018}; \cite{Steinhardt:2002}, \cite{steinhardt_2007}; \cite{IjjasSteinhardt:2017}), there could have been a single previous universe that underwent a contraction phase that `bounced' into our present universe (for reviews of models in the previous two categories, see \cite{Kragh_2009}; \cite{Kragh_2018}; \cite{Lilley:2015ksa}; \cite{Novello:2008ra}; \cite{Brandenberger2017}), or there could be a multiverse in which the entire multiverse contracted and subsequently `bounced' (\cite{Aguirre:2002}; \cite{Aguirre:2003}; \cite{Carroll:2004}). I will refer to cosmologies in all three categories as `bounce cosmologies'. As I've noted, one reason the BGV theorem does not apply to bounce cosmologies involves the postulation that the universe's average expansion rate is less than or equal to zero. However, the BGV theorem is inapplicable to many bounce cosmologies for the additional (and arguably more important) reason that they postulate a non-classical (quantum) regime to which theorems about classical space-times do not apply; see, for example, \cite{Huggett:2018}.

Previously, we saw that ALP postulates the Past Hypothesis, that is, the hypothesis that the beginning of our universe was a stupendously low entropy state. Consistent with the Past Hypothesis, bounce cosmologies postulate that the entropy reached a minimum at the interface between universes (or, in the multiverse models, that the multiverse reached an entropy minimum on an interface between two stages of the multiverse). Craig and Sinclair have argued that the orthodox interpretation of bounce cosmologies is mistaken. On their view, the entropic arrow of time aligns with the direction of time. Bounce cosmologies postulate an interface such that the arrow of time points away from the interface in two directions. So, in Craig and Sinclair's interpretation, the two directions that the entropic arrow points away in must both lie in the future of the interface. As they write, `The boundary that formerly represented the ``bounce'' will now [be interpreted to] bisect two symmetric, expanding universes on either side' \cite[p. 122]{CraigSinclair:2012}. Elsewhere, they conclude from this feature that, `The last gambit [in trying to avoid an absolute beginning], that of claiming that time reverses its arrow prior to the Big Bang, fails because the other side of the Big Bang is \emph{not} the past of our universe' \cite[p. 158]{CraigSinclair:2009}. As they continue, `Thus, the [universe on the other side of the interface] is not our past. This is just a case of a double Big Bang. Hence, the universe \emph{still} has an origin' \cite[pp. 180-1]{CraigSinclair:2009}; also see \cite[pp. 125-7]{CraigSinclair:2012}, \cite[pp. 61, 77-8]{CarrollCraig:2016}.\footnote{A large proportion of the argumentation that that Craig and Sinclair offer in their ([\citeyear{CraigSinclair:2009}], [\citeyear{CraigSinclair:2012}]) concerns the Aguirre-Gratton model ([\citeyear{Aguirre:2002}, [\citeyear{Aguirre:2003}]. Nonetheless, Craig and Sinclair intend for their conclusions to generally apply to cosmological model in which the entropic arrow reverses direction at an interface (for example, \cite[p. 158]{CraigSinclair:2009}). For that reason, Craig takes the conclusion to apply to the Carroll-Chen model in his debate with Carroll ([\citeyear{CarrollCraig:2016}]) and to apply to Penrose's Conformal Cyclic Cosmology ([\citeyear{CraigOnPenrose}]).} In other words, according to Craig and Sinclair, bounce cosmologies describe an absolute beginning after all. And -- so Craig and Sinclair argue -- this absolute beginning requires an efficient cause beyond either of the two universes (or beyond the multiverse).\footnote{Other authors have provided arguments for the conclusion that bounce cosmologies should be interpreted as depicting the birth of two universes instead of the death of one universe and the birth of another. For example, Nick Huggett and Christian W{\"u}thrich have argued that string cosmological models and loop quantum cosmological models depict two universes being born from a non-spatio-temporal regime \cite{Huggett:2018}. Nonetheless, Huggett and W{\"u}thrich's interpretation cannot help Craig and Sinclair; Huggett and W{\"u}thrich's interpretation explicitly depends on the reducibility of space-time to non-spatiotemporal phenomena. If time is reducible then we are back at the first arm of the Interface Dilemma.}

Nonetheless, Craig and Sinclair have argued that the direction of time is not reducible. If the direction of time is not reducible, then we are left without reason to think that the direction of time aligns with the entropy gradient. Consequently, if the direction of time is not reducible, then the direction of time need not point away from the interface in two directions. So, Craig and Sinclair's interpretation of the interface as an absolute beginning is unjustified. Moreover, on Craig and Sinclair's anti-reductionism, there's no longer reason to suppose that the direction of efficient causation would align with the entropy gradient. As I will argue in section \ref{interp_section}, events in a cosmological epoch of higher entropy could be the causes of events in an epoch of lower entropy or the universes to either side of the interface might be interpreted as the simultaneous causes of each other. So, we're left without reason to deny that events on one side of the interface could be the efficient causes of events on the other side of the interface.

\subsection{An objection to premise \ref{if_not_reducible_then_not_cor}}

Like the first arm, the second arm of the dilemma is a valid argument. In this case, objections can plausibly be made only to premises \ref{if_not_reducible_then_not_cor} and \ref{premiseAandB}. According to premise \ref{if_not_reducible_then_not_cor}, if  the  direction of time  is  not  reducible  then  the direction  of  time  does  not  necessarily  correspond  to  the  direction  of  the  entropic arrow of time. One might think that this premise is obvious; if the temporal directedness of macrophysical processes and the entropic arrow of time are independent, then there is no need for them to align. However, according to Craig and Sinclair, friends of tensed theories of time have reason to think that the direction of time and the entropic arrow do align. Craig and Sinclair argue that the alignment of `the cosmological, thermodynamic, electromagnetic, and psychological arrows of time' can be explained by postulating that each of the arrows are the `physical manifestations of the same underlying temporal becoming of metaphysical time' \cite[p. 137]{CraigSinclair:2012}. 

The notion of a `physical manifestation' of absolute (or `metaphysical') time is obscure; neither Craig nor Sinclair have provided an analysis of what that amounts to. Craig has suggested that the entropic arrow aligns with the direction of time as a matter of nomic necessity \cite[p. 78]{CarrollCraig:2016} and perhaps this is one analysis of `physical manifestation'. Craig takes the alignment between the direction of time and the entropic arrow to be nomologically necessary because of the second law of thermodynamics. The second law of thermodynamics states that, for closed systems, the entropy increases towards the future. We can understand the universe (or multiverse) as a closed system so that we would expect the total entropy to increase into the future. But we already know that the second law of thermodynamics is a statistical regularity that admits of exceptions.

For example, consider a world consisting of a vast collection of Newtonian particles at thermodynamic equilibrium. Because the world is at equilibrium, the average entropy of the world will have some constant value.  Nonetheless, the entropy will fluctuate around that average, and there will be some finite probability for fluctuations of arbitrary size. Given enough time, the system will fluctuate to a microscopically small entropy. To either side of the entropy minimum, the direction of the entropic arrow of time will point away  from the time at which the entropy is minimized. For ALP, time could be interpreted as `flowing' away from the minimum in either direction. Thus, ALP will not be able to maintain, in any realistic sense, that the system was first at equilibrium, then fluctuated into an entropy minimum, and, lastly, fluctuated back to some higher entropy state. To their advantage, authors who endorse the irreducibility of temporal becoming will be able to say that these three events occurred in the order indicated regardless of the direction of the entropic arrow. If friends of tensed theories of time maintain this interpretation of equilibrium systems, then they would likewise have no reason to suppose that the interface is the beginning of two universes as opposed to the death of one universe and the birth of another. In fact, Craig has provided an argument that strongly suggests this interpretation. He writes:

\begin{quote}
    A great deal of ink has been spilled in the attempt to ground time's asymmetry in various physical processes such as entropy increase, the expansion of the universe, and so forth. From a theistic perspective, however, all such attempts seem misconceived. For one can easily conceive of a possible world in which God creates a universe lacking any of the typical thermodynamic, cosmological or other arrows of time, and yet He experiences the successive states of the universe in accord with the lapse of His absolute time \cite[p. 162]{craig_2001}.
\end{quote}

One can likewise imagine God creating a universe in which God experiences the successive states of absolute time passing through states of decreasing entropy. This is one way the world could lack `any of the typical thermodynamic, cosmological or other arrows of time'. A parallel argument is open to atheistic friends of absolute time. Provided the existence of absolute time, in a world sans God, one can easily imagine a cosmic perspective, not occupied by any agent, in which absolute time passes through states of decreasing entropy. For as Craig notes (emphasis his), `such physical processes [for example, entropy increase] are simply irrelevant to a definition of temporal asymmetry. For why should we regard one direction of the physical process as the ``earlier'' direction rather than the ``later'' direction?' \cite[p. 162]{craig_2001} On the account that Craig endorses, we may have empirically discovered that the passage of time is correlated with entropy increase, but this correlation does not reflect anything deep about the direction of time, and situations can arise in which this correlation becomes broken. Consequently, provided that Craig's theory of absolute time is correct, the alignment between the direction of time and the entropic arrow is not logically, metaphysically, or nomologically necessary.

The Mentaculus project has already provided a promising dynamical explanation of the alignment of the various arrows. In order to provide an alternative explanation of the alignment of the various arrows of time in terms of the flow of absolute time, Craig would need to posit that the arrows are not independent of the flow of absolute time. Nonetheless, Craig's conception of the objective becoming of absolute time differs from conceptions of tensed facts as provided by more naturalistically inclined philosophers -- such as Tim Maudlin -- because Craig's absolute time is supposed to be independent of the physical world. One has difficulty seeing how Craig's absolute time could matter for the arrows' directions unless the dynamics of physical objects were appropriately linked to absolute time.\footnote{See the related objections that  W{\"u}thrich has provided to Craig \cite[pp. 264-5]{Wuthrich:2010} and that Loewer has provided to Maudlin \cite[pp. 133-5]{Loewer:2012b}. Also see \cite{Prosser:2000}; \cite{Miller:2017}.} Perhaps future physics will reveal some appropriate way to incorporate the objective becoming of absolute time. Various authors have tried to postulate something that approaches a dynamical role for objective becoming that would tie objective becoming to the entropic arrow. But their research programmes have yet to succeed in showing an appropriate link between objective becoming and the arrows. For example, though not a friend of tensed theories of time, Weingard postulates a role for a time-ordering field, that is, a vector field that plays the role of determining the direction of time \cite{Weingard_1977}. But in order to perform this dynamical role, the field would need to couple to matter. Craig Callender calls the view `interesting yet embryonic' because Weingard does not tell us how such a field could couple to matter \cite{Callender_2016}. The problem is more severe than Callender supposes because no field has been detected, yet we have reason to think we should be able to detect the time-ordering field if the field exists. If another field coupled to (for example) electromagnetism and the coupling were strong enough that the field's effects would be detectable in ordinary experience, we should be able to perform a high energy experiment that produced quanta of that field \cite[pp. 180-3]{Carroll_2016}. Yet we have never observed quanta of a time-ordering field.

Perhaps the violation of $CP$-symmetry could (somehow) provide something approaching a dynamical role for the order of objective becoming. \cite{Maudlin_2002} and \cite[p. 16]{albert_2000} have endorsed the view that $CP$-symmetry violation is temporally asymmetric. On this point, John Earman has provided a succinct discussion that is worth reflecting on \cite[pp. 258-9]{Earman_2002}. Earman writes that, `For if what has been experimentally detected is in fact a failure of time reversal invariance of the fundamental laws of physics —- and if these laws do have a universal character -— then we can be sure that our universe is temporally orientable and that it is in fact temporally oriented' \cite[p. 259]{Earman_2002}. Nonetheless, supposing that the fundamental physical laws do tell us that time has some orientation, the laws are silent on which orientation is the correct one. Earman writes, 

\begin{quote}
    For a lawlike equation of motion $L$ that is not time reversal invariant will be matched by a lawlike $^TL$ that is also not time reversal invariant and that makes exactly the opposite division of histories into those that are allowed and those that are not. So unless $L$ and $^TL$ are distinguished by time orientation neutral predictions, the failure of time reversal invariance tells us that there is a time orientation but not which is the correct one. As far as I am aware, there are no plausible candidates for the role of what might be called time reversal super-non-invariant laws $SL$ having the property that not only are $SL$ not time reversal invariant but also that $ST$ and $^TSL$ are distinguished by time orientation neutral predictions \cite[p. 259]{Earman_2002}.
\end{quote}

\subsection{An objection to premise \ref{premiseAandB}}

Recall that premise \ref{premiseAandB} states that if the direction of time does not necessarily correspond to the direction of the entropic arrow of time then (a) the interpretation of the interface as the beginning without a physical cause is unjustified and (b) the claim that events on one side of the interface cannot be the efficient causes of events on the other side of the interface is unjustified. To object to this premise, one would need to provide a compelling reason for thinking that the antecedent could be true when the consequent is false. That is, one would need to show that while the direction of time does not necessarily correspond to the direction of the entropic arrow of time, the interpretation of the interface as a beginning without a physical cause is justified or the claim that events on one side of the interface cannot be the efficient causes of events on the other side of the interface is justified.

Craig and Sinclair think they can provide such a justification. To unpack their argument, let's put some more machinery on the table. Two space-time points are causally connected if there exists a physically realizable trajectory connecting the two points. Trajectories are physically realizable if a particle can travel along the trajectory without exceeding the speed of light. Causal connection is, at best, a physically necessary, but not sufficient, condition for efficient causation. Two events can be causally connected even though neither event was the efficient cause of the other event. For example, the American Civil War was causally connected to the assassination of the archduke Franz Ferdinand, but the Civil War was not the efficient cause of Ferdinand's assassination. Moreover, if efficient causation is reducible, so that microphysical events do not have efficient causes, then two microphysical events can still be causally connected. Causal connection is a symmetric relation. However, the ability to send a macrophysical signal from $A$ to $B$ is asymmetric; macrophysically, if a signal is transmitted from $A$ and received at $B$, then a signal cannot be transmitted from $B$ and received at $A$. There is a radiation arrow of time that determines the directions in which a macrophysical signal can be allowably transmitted or received.

To support their interpretation of the interface as a double Big Bang, Craig and Sinclair describe how, at least in Aguirre and Gratton's model [\citeyear{Aguirre:2002}], an observer on one side of the interface cannot transmit to or receive from an observer on the other side of the interface for two reasons \cite[123]{CraigSinclair:2012}. First, Craig and Sinclair take the fact that the interface in the Aguirre-Gratton model is a null surface to entail that the regions on either side of the interface are causally disconnected from each other. Second, because the radiation arrow points away from the interface, we might think we should interpret the interface as located in the past for observers on either side of the interface. Thus, we would have reason -- independent of the entropic arrow -- for thinking that the interface is the absolute beginning for two universes.

This objection does not succeed. On the one hand, Aguirre and Gratton do not regard their model as physically realistic. Instead, their model is meant to mathematically illustrate one example of an expanding universe to which the BGV theorem does not apply. As Aguirre and Gratton themselves admit, their example `is not a viable cosmological model' [\citeyear[p. 4]{Aguirre:2002}]. Indeed, even after developing a more sophisticated version of their model in their [\citeyear{Aguirre:2003}], Aguirre and Gratton state that their model will need to be more fully developed in a quantum gravitational context in order to serve as a launching point for a realistic cosmology. Their model is meant to illustrate principles which have relevance for physically realistic models. In one of the physically realistic models that Aguirre and Gratton consider, there are nomologically allowed trajectories that do traverse the interface \cite[p. 4]{Aguirre:2002}. Thus, the causal disconnection between universes on either side of the interface is not a generic feature of bounce cosmologies. Still, one could worry that the interface should be interpreted as an absolute beginning insofar as the radiation arrow points away from the interface, even if the two universes are not causally disconnected. Thus, the radiation arrow might still provide us with a justification, independent of the entropic arrow, for thinking that the interface is an absolute beginning.

Friends of ALP argue that the Mentaculus offers a reductive explanation of the radiation arrow. If that reduction is successful, we have the first arm of the Interface Dilemma. On the other hand, if that reduction is not successful then we can restate the Interface Dilemma by modifying premises \ref{if_not_reducible_then_not_cor} and \ref{premiseAandB}. In the absence of reduction, we had no reason for thinking that the entropic arrow and the direction of time necessarily align. Likewise, we would have no reason for thinking the radiation arrow and the direction of time necessarily align. Without reason to think that the direction of time and the radiation arrow necessarily align, we would have no reason to suppose that the interface is in the past for observers in both universes or, despite the inability to transmit signals through the interface, that the goings-on on one side of the interface cannot be the efficient causes of goings-on on the other side of the interface.

\subsection{\label{interp_section}How should friends of absolute time interpret bounce cosmologies?}

For Craig and Sinclair, the alignment between the entropic arrow and the direction of time provides friends of tensed theories of time with additional reason to favor the view that the interface is an absolute beginning of two universes. To the contrary, friends of tensed theories of time have additional reason to reject Craig and Sinclair's interpretation. As I will proceed  to explain, there are two possibilities. On the one hand, tensed theories of time might provide reason to interpret the interface as a beginning, but in a sense incompatible with the theistic implications of the \emph{Kal$\overline{\textrm{a}}$m} argument. In this case, I will argue that the two universes can be the simultaneous causes of each other. Alternatively, tensed theories of time provide reason to interpret the interface as a transition from one universe to another.\footnote{Elsewhere, I argue that there are bounce cosmologies in which there is an asymmetric explanatory relationship between the universes on either side of the interface so that one of the two universes must precede the other in time [Under review -- redacted]. I will suppose that the arguments that I offer there do not suffice to show that one universe must precede the other in time. Alternatively, the arguments that I offer in this section can be understood to be restricted to those bounce cosmologies to which my arguments in [redacted] are inapplicable.} In this case, we wouldn't have a reason to postulate a beginning of the universe.

Let's suppose tensed theories do provide us some reason to interpret the interface as a beginning. Let's call the two universes, on either side of the interface, $U_1$ and $U_2$. The interface is a feature of both $U_1$ and $U_2$. Therefore, the interface can be interpreted as a feature of $U_1$ that causes the beginning of $U_2$ by being simultaneous with the beginning of $U_2$. One might then ask what causes $U_1$ to begin to exist. Given the symmetry between $U_1$ and $U_2$, $U_2$ can be interpreted as the simultaneous cause of $U_1$. In other words, $U_1$ and $U_2$ can be interpreted as the simultaneous causes of each other. Given that the complex  of two universes can be understood as a closed system of causes, the need to explain the complex in terms of some external supernatural cause does not arise.

The interpretation that the interface represents two universes simultaneously causing each other might seem implausible because, on presentism, at the first instant of time, only the interface exists. Consequently, we might worry that Craig and Sinclair would object that the interface requires a cause. However, this reply is not available to Craig. On Craig's view, the present is not a temporal interval with zero duration. In fact, Craig has argued that temporal intervals with zero duration -- what might be called `instants' -- do not exist at all. He writes that, `it seems to me very difficult to reconcile the A-theory of time [which Craig endorses] with the view that instants [...] subsist as independent, degenerate intervals of zero duration.' Craig explains that, on his view, `only intervals of time are real or present and that the present interval (of arbitrarily designated length) may be such that there is no such time as ``the present'' \emph{simpliciter}; it is always ``the present hour'', ``the present second'', etc. The process of division is potentially infinite and never arrives at instants' \cite[p. 260]{Craig_1993_Criticism}. Craig specifies that instead of holding an atomistic conception of time, he maintains a view on which `only intervals of time are real or present and that the present interval may be subdivided into subintervals which are past, present, and future respectively' \cite[p. 260]{Craig_1993_Criticism}. Elsewhere, Craig writes, `[...] anything having positive ontological status would seem necessarily to exist for some temporal duration; to say it exists only at a durationless instant is to ascribe reality to a mathematical chimera' \cite[p. 499]{Craig:1991}.\footnote{One might add that Craig presents two lines of defense for the second premise of the \emph{Kal$\overline{\textrm{a}}$m} argument. Craig's first line of defense for the second premise of the \emph{Kal$\overline{\textrm{a}}$m} argument commits him to denying the existence of instants. In Craig's first line of defense, Craig offers a number of a priori arguments for the conclusion that there are no actual infinities. If there are no actual infinities, then the past series of events is not infinite. Provided that each event in the past series has a finite duration, Craig concludes that the universe must have existed only for a finite interval of time. An actual infinity of anything would be inconsistent with Craig's a priori argument. But, if instants exist, any finite interval of time would contain an actual infinity of instants.}

Supposing that there are no instants, there is no instant at which only the interface exists. Instead, there exists an interval of time whose boundary is the interface between the two universes. Consequently, on Craig and Sinclair's interpretation of bounce cosmologies, for every existent temporal interval, $U_1$ and $U_2$ co-exist. There is no need to introduce an independent cause for the interface, and we can interpret $U_1$ and $U_2$ as the simultaneous causes of each other. 

Here, another objection to the interpretation I've offered might occur to the reader. If the two universes share only a boundary, and boundaries should (according to Craig) be interpreted as mere mathematical fictions to which we should not be ontologically committed, then we should not be ontologically committed to the interface. And if we should not be ontologically committed to the interface, then we should not be ontologically committed to the only place in the model where a relation of efficient causation could be thought to obtain between the two universes. If we are not ontologically committed to the only place in the model where a relation of efficient causation could obtain between the two universes, then we should not be committed to both universes being the efficient causes of each other. 

This objection is undermined by a puzzle concerning the extendibility of geodesics through the interface. In singular cosmologies, the Big Bang singularity is interpreted as a past boundary to space-time. A space-time in which geodesics cannot be extended past some point or region are naturally interpreted as space-times containing a boundary. Bounce cosmologies are typically non-singular; importantly, so long as the two universes are not causally disconnected, geodesics can be extended through the interface. In the orthodox interpretation, the extendibility of a geodesic through the interface suggests that a particle could travel along a trajectory from one universe, through the interface, and into the other universe. But on the interpretation that we are now considering, the extendibility of geodesics through the interface cannot be interpreted in terms of a particle traveling through the end of one universe and into the beginning of another universe. For that reason, Craig and Sinclair favor an instrumentalist interpretation of the extendibility of geodesics through the interface. But recall that, for Craig, instants are mathematical fictions; we should talk about, for example, a present interval and not the present \emph{simpliciter}. In orthodox interpretations of General Relativity, we imagine that a particle traveling along a given geodesic carries a clock that parametrizes the trajectory in terms of the proper time. On this interpretation, there exists an instant for each proper time along the geodesic. At one proper time, the geodesic pierces the interface. But Craig understands instants -- and so specific proper times -- as mathematical fictions. So, Craig and Sinclair should say that there exist intervals of time (that is, all of those intervals which contain the interface) in which features of one universe are continuous into the other universe.

Consider the relativity of simultaneity. Observer $A$ can receive signals $S_1$ and $S_2$ from two causally disconnected events $E_1$ and $E_2$. For $A$, $E_1$ and $E_2$ appear simultaneous. Another observer $B$, traveling at some velocity close to the speed of light relative to $A$, might observe $E_1$ occurring before $E_2$. We can suppose that $A$ and $B$ travel along geodesics that pierce the interface. So, $A$ and $B$ may disagree about whether $E_1$ or $E_2$ occur before, at, or after the interface. On interpretations of relativity without absolute time, there is no objective fact about which space-like surface is the interface; one can foliate space-time into space-like surfaces in multiple ways, so that what is considered part of $U_1$, $U_2$, or the interface will depend upon which foliation one chooses. In that case, there is no objective fact as to whether $A$ or $B$ observed the correct ordering of $E_1$ and $E_2$. 

According to interpretations of relativity that include absolute time, there is an absolute relation of simultaneity, such that which events are on the interface or in either universe is not the result of one's choice of foliation; on an interpretation like that, there is an objective fact as to whether $E_1$ occurred before $E_2$. Granting that such a relation can be sensibly constructed, there will be some objective fact about whether $E_1$ or $E_2$ occurred in $U_1$, $U_2$, or on the interface. But this is beside the point. We can engineer the thought experiment so that $E_1$ objectively occurred in $U_1$ and $E_2$ objectively in $U_2$. Despite this, while in $U_1$, $A$ observes $E_1$ and $E_2$ occurring simultaneous with the interface. Even if this appearance is illusory, the content of the illusion can be explained only by invoking features of both universes.\footnote{Readers might object that if the radiation arrow is directed away from the interface, then an observer could not make intelligible observations using signals from both universes. But this is beside the point. My presentation of the thought experiment may have been in terms of observers, but the thought experiment was meant to drive intuitions concerned the sort of formal relations we can construct on the space-time of interest. Even if no agent could rationally construct the observations required for the thought experiment, the requisite formal relations that the thought experiment was meant to illustrate can still be constructed.} That is, Craig and Sinclair would have to say that there exists an interval shared by both universes in order to explain features of the interface.

Be that as it may, friends of metaphysical time have at least as much -- if not more -- reason to adopt the view that the interface is a transition between one universe and another and not the absolute beginning of two universes. In order to interpret the interface as an absolute beginning to two universes, Craig and Sinclair need to reinterpret extendibility as `a technical artefact rather than an indication of past eternality' \cite[p. 127]{CraigSinclair:2012}. Those of us who are committed to the claim that theories in physical cosmology aim for approximate truth, and not mere empirical adequacy, should not endorse an instrumentalist reading of an aspect of a given cosmological model unless we have reason to do so.\footnote{Craig and Sinclair should generally be committed to the the realist reading because they would like to draw metaphysical conclusions from cosmological models. In any case, I'm not claiming that any bounce cosmological model is approximately true. Instead, I am only committing myself to the thesis that cosmological models attempt to describe the world and so should be read realistically, even if what they propose about the world turns out to be false. In contrast, an instrumentalist reading of bounce cosmologies would maintain that bounce cosmologies do not aim to represent the approximate truth about our universe. Instead, bounce cosmologies aim only at, for example, empirical adequacy. In that case, even if we came to endorse a bounce cosmology, we need not become ontologically committed to the unobservable consequences of that cosmology.} If there is a realist reading of the extendibility of geodesics through the interface that is consistent with the reality of the objective becoming of metaphysical time, then, all else being equal, friends of tensed theories will have reason to endorse the realist reading. Since friends of tensed theories need not endorse the view that the direction of temporal becoming and the entropic arrow of time align, the fact that geodesics can be extended through the interface provides some reason to think that the interface should be interpreted as a transition from one universe to another and not as an absolute beginning for two universes. Again, we can imagine God -- or the fictional agent occupying the view-from-nowhere -- watching as metaphysical time passes, the world unfurling through ages of entropy decrease, until the entropy begins to increase once more.

\section{Conclusion}

We can add a tautologous premise and assemble the previous two subarguments in order to present the Interface Dilemma in full:

\begin{enumerate}
    \item If the direction of time is reducible then, probably, not all events have efficient causes. 
    \item If, probably, not all events have efficient causes then the first premise of the \emph{Kal$\overline{\textrm{a}}$m} argument is probably false. 
    \item So, if the direction of time is reducible then the first premise of the \emph{Kal$\overline{\textrm{a}}$m} argument is probably false (hypothetical syllogism from \ref{if_reducible_then_not_all}, \ref{if_not_all_then_P1_false}). 
    \item If the direction of time is not reducible then the direction of time does not necessarily correspond to the direction of the entropic arrow of time. 
    \item If the direction of time does not necessarily correspond to the direction of the entropic arrow of time then
    \begin{enumerate}
        \item the interpretation of the interface as a beginning without a physical cause is unjustified and 
        \item the claim that events on one side of the interface cannot be the efficient causes of events on the other side of the interface is unjustified. 
    \end{enumerate}
    \item If (\ref{beginning_unjustified}) and (\ref{eff_cause_unjustified}) then Craig and Sinclair's interpretation of bounce cosmologies is unjustified. 
    \item So, if the direction of time is reducible then Craig and Sinclair's interpretation of bounce cosmologies is unjustified (hypothetical syllogism from \ref{if_not_reducible_then_not_cor}--\ref{If_ab_then_unjustified}). 
    \item Either the direction of time is reducible or not (tautology).\label{reducible_or_non-reducible} 
    \item Therefore, either the first premise of the \emph{Kal$\overline{\textrm{a}}$m} argument is probably false or Craig and Sinclair's interpretation of bounce cosmologies is unjustified (constructive dilemma from \ref{subconc1}, \ref{subconc2}, and \ref{reducible_or_non-reducible}). 
\end{enumerate}

The Interface Dilemma is a valid argument. A successful objection must show either that at least one of the premises has not been adequately supported, that one of the premises is unlikely to be true, or that, even though each premise is more probable than not, the conjunction of the premises is improbable. However, as I described, objections can be plausibly constructed only for premises \ref{if_not_all_then_P1_false}, \ref{if_not_reducible_then_not_cor}, and \ref{premiseAandB}. We've seen that a variety of objections to those premises fail.

\section*{Acknowledgements}

I want to thank Elizabeth Jelinek for kindly hosting me while I prepared this manuscript. In addition, I'd like to thank Levi Greenwood and Martin Curd for long, invaluable conversations, as well as anonymous reviewers at \emph{The British Journal for Philosophy of Science} for their feedback.

\begin{flushright}
\emph{
  Dan Linford\\
  Purdue University\\
  Department of Philosophy\\
  100 N University St\\
  West Lafayette, IN 47907\\
  dlinford@purdue.edu
}
\end{flushright}

\pagebreak

\bibliographystyle{plainnat}
\bibliography{references.bib}{}

\end{document}